\documentclass{elsart}

\usepackage{amsfonts}
\usepackage{amsmath}
\usepackage[dvips]{graphicx}
\usepackage{psfrag}

\newcommand{\heta}{\hat{\eta}}
\newcommand{\hu}{\hat{u}}
\newcommand{\hv}{\hat{v}}
\newcommand{\hw}{\hat{w}}
\newcommand{\sD}{\mathcal{D}}
\newcommand{\rme}{\mathrm{e}}
\newcommand{\vect}[1]{\mathbf{#1}}

\begin{document}

\begin{frontmatter}

\journal{Journal of Computational Physics}

\title{A low-cost parallel implementation of direct numerical
simulation of wall turbulence}

\author[aut1]{Paolo Luchini\corauthref{corr1}},
\ead{luchini@unisa.it}
\author[aut2]{Maurizio Quadrio}
\ead{maurizio.quadrio@polimi.it}

\address[aut1]{Dipartimento di Ingegneria Meccanica Universit\`a di 
Salerno - Italy}

\address[aut2]{Dipartimento di Ingegneria Aerospaziale Politecnico 
di Milano - Italy}
                                                
\corauth[corr1]{Corresponding author. Address: Dipartimento di
Ingegneria Meccanica Universit\`a di Salerno - via Ponte don Melillo -
84084 Fisciano (SA) - Italy}

\begin{abstract}
A numerical method for the direct numerical simulation of
incompressible wall turbulence in rectangular and cylindrical
geometries is presented. The distinctive feature resides in its design
being targeted towards an efficient distributed-memory parallel
computing on commodity hardware. The adopted discretization is
spectral in the two homogeneous directions; fourth-order accurate,
compact finite-difference schemes over a variable-spacing mesh in the
wall-normal direction are key to our parallel implementation. The
parallel algorithm is designed in such a way as to minimize data
exchange among the computing machines, and in particular to avoid
taking a global transpose of the data during the pseudo-spectral
evaluation of the non-linear terms. The computing machines can then be
connected to each other through low-cost network devices. The code is
optimized for memory requirements, which can moreover be subdivided
among the computing nodes. The layout of a simple, dedicated and
optimized computing system based on commodity hardware is
described. The performance of the numerical method on this computing
system is evaluated and compared with that of other codes described in
the literature, as well as with that of the same code implementing a
commonly employed strategy for the pseudo-spectral calculation.
\end{abstract}


\end{frontmatter}

\section{Introduction}

The direct numerical simulation (DNS) of the Navier--Stokes equations
written for low-Reynolds-number, incompressible turbulent flows in
simple geometries is becoming an increasingly valuable tool for basic
turbulence research
\cite{moin-mahesh-1998}. Interesting wall-bounded flows span a number
of simple geometries, either in cartesian (plane channel flow,
boundary layer over a flat plate) or in cylindrical (pipe flow,
annular pipe flow, flow in curved channels) coordinate systems.

For the cartesian case, an effective formulation of the equations of
motion was presented 15 years ago by Kim, Moin \& Moser in their
pioneering and widely-referenced work on the DNS of turbulent
plane-channel flow \cite{kim-moin-moser-1987}. This formulation can be
regarded today as a {\em de facto} standard; it has since then been
employed in many of the DNS of turbulent wall flows in planar
geometries. It consists in the replacement of the continuity and
momentum equations written in primitive variables by two scalar
equations, one (second-order) for the normal component of vorticity
and one (fourth-order) for the normal component of velocity, much in
the same way as the Orr--Sommerfeld and Squire decomposition of linear
stability problems. The main advantages of such an approach are that
pressure is eliminated from the equations, and the two wall-parallel
velocity components are recovered as the solution of a $2 \times 2$
algebraic system (a cheap procedure from a computational point of
view), when a Fourier expansion is adopted for the homogeneous
directions. A high computational efficiency can thus be achieved. The
same approach can be employed to write the equation in cylindrical
coordinates, but it appears to be much less popular than the
primitive-variable formulation of the Navier--Stokes equations. The
formulation in terms of two scalar equations for radial velocity and
radial vorticity can be found for example in
\cite{quadrio-luchini-2002}.

This optimally efficient formulation does not prescribe any particular
discretization method for the differential operators in the
wall-normal direction. Many researchers, including Kim {\em et al.}
\cite{kim-moin-moser-1987}, used spectral methods (typically Chebyshev
polynomials) in this direction too, but other possibilities exist,
finite differences and B-splines \cite{kwok-moser-jimenez-2001} being
the most popular ones. The use of finite differences has seen growing
popularity
\cite{moin-mahesh-1998}, but mainly in the context of the 
primitive-variable formulation, and for the discretization of the
derivatives in {\em all} three spatial directions (see for example
\cite{na-moin-1998}).  The choice of spectral methods for the
discretization of the wall-normal, inhomogeneous direction has a
direct impact on the parallelization of a computer code, given the
non-locality of the spectral differential operators. As a matter of
fact, to our knowledge no fully spectral DNS code has been able to
date to run in parallel without a large amount of communication. As a
consequence, high-performance parallel DNS has been mostly restricted
to large computing systems with a specially-designed communication
infrastructure, a.k.a. supercomputers, even though the floating-point
computing performance of the modern, mass-marketed CPUs is comparable
or better than those of supercomputers
\cite{dongarra-2004}. In a recent paper 
\cite{jimenez-2003} Jim\'enez draws an interesting picture of 
the future of the DNS of turbulent flows in the next 30 years, and
assumes that such simulations will be run on supercomputers. The work
by Karniadakis and coworkers \cite{karamanos-etal-1999} makes no
exception, in that it shows that DNS of turbulent flows can be carried
out with reasonably good performance on a cluster of PC, provided they
are interconnected by a high-performance Myrinet network. When a
Beowulf cluster of PC connected with standard Fast Ethernet cards is
employed \cite{dmitruk-etal-2001} on a isotropic turbulence problem,
even after extensive optimization of the code the parallel efficiency,
i.e. the ratio between total time on one processor and $p$ times the
computing time on $p$ processors, is as low as 0.5 already when two
machines are used.

In this paper we present a numerical method for the DNS of turbulent
wall flows that has been designed to require a limited amount of
communication, and thereby is well suited for running on commodity
hardware. The method is based on the standard normal velocity - normal
vorticity formulation and hence uses Fourier discretization in the
homogeneous directions, but high-order, compact finite differences
schemes are chosen for the discretization of the wall-normal
direction, instead of the classical expansion in terms of Chebyshev
polynomials. It can be used with either cartesian or cylindrical
coordinates.

The outline of the paper is as follows. In
\S\ref{sec:equations} the cartesian form of the Navier--Stokes
equations which is best suited for their numerical solution, and their
Fourier discretization with respect to the homogeneous directions are
briefly recalled. In \S\ref{sec:timescheme} time discretization is
discussed, with emphasis on our strategy for data storage, which
allows us to achieve an important memory optimization.  In
\S\ref{sec:cart-fd} the finite-difference discretization of the
wall-normal direction based on explicit compact schemes is
introduced. \S\ref{sec:parallel} illustrates the parallel algorithm,
which takes advantage of distributed-memory
(\S\ref{sec:distributed-mem}) as well as shared-memory
(\S\ref{sec:shared-mem}) machines;
\S\ref{sec:personal-supercomputer} describes how a specialized
parallel computing system can be set up to achieve the highest
efficiency based on commodity hardware.
\S\ref{sec:performance} discusses the
performance of the present parallel method, compared to the
information available in the literature for similar codes. A
comparison with a classical parallel strategy often employed in
similar codes is used in \S\ref{sec:transpose} to assess the
usefulness of the present method when used with low-bandwidth network
connections. Lastly, \S\ref{sec:conclusions} is devoted to
conclusions.

\section{The numerical method}
\label{sec:label}

\subsection{Governing equations and spectral discretization}
\label{sec:equations}

Here we describe only the main aspects of the numerical method. Full
details, including the extension to the cylindrical case, can be found
in \cite{quadrio-luchini-2004}.

\begin{figure}
\centering
\psfrag{Lz}{$L_z$}
\psfrag{Lx}{$L_x$}
\psfrag{yu}{$y_u$}
\psfrag{yl}{$y_l$}
\psfrag{flow}{$flow$}
\psfrag{2d}{$2 \delta$}
\psfrag{xu}{$x,u$}
\psfrag{yv}{$y,v$}
\psfrag{zw}{$z,w$}
\includegraphics[scale=0.5]{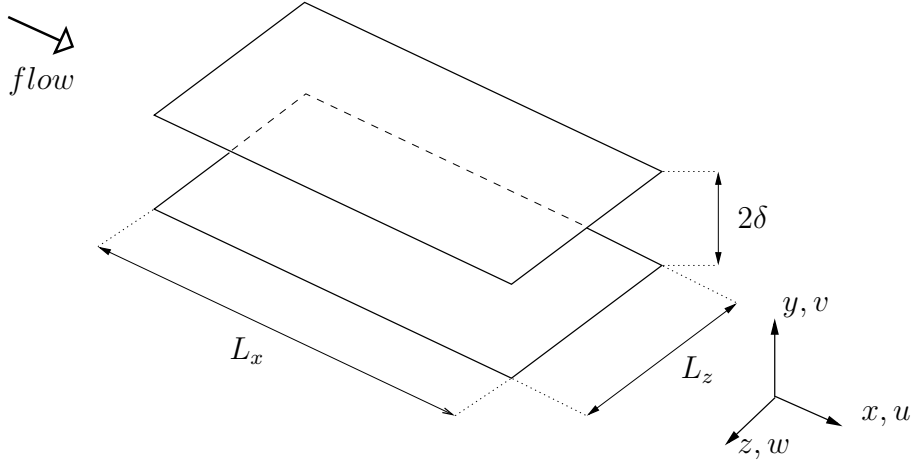}
\caption{Sketch of the computational domain for the cartesian 
coordinate system.}
\label{fig:cart-comp-domain}
\end{figure}

Our cartesian coordinate system is illustrated in figure
\ref{fig:cart-comp-domain}, where a sketch of a plane channel flow is 
shown: $x$, $y$ and $z$ denote the streamwise, wall-normal and
spanwise coordinates, and $u$, $v$ and $w$ the respective components
of the velocity vector. The flow is assumed to be periodic in the
streamwise and spanwise directions. The reference length $\delta$ is
taken to be one half of the channel height. 

The non-dimensional Navier--Stokes equations for an incompressible
fluid in cartesian coordinates are rewritten, following
\cite{kim-moin-moser-1987}, in terms of two scalar
differential equations, one (second order) for the wall-normal
component of vorticity $\eta$ and one (fourth-order) for the
wall-normal component of velocity $v$, and then Fourier-transformed
along the homogeneous directions (Fourier-transformed variables will
be indicated with an hat sign). If the nonlinear terms are considered
to be known, as is the case when such terms are treated explicitly in
the time discretization, these equations (supplemented by no-slip
boundary conditions at the walls) become uncoupled and can be solved
separately to advance the solution in time by one step. Computing the
nonlinear terms and their spatial derivatives requires us first to
compute $\hu$ and $\hw$ from $\hv$ and $\heta$.  By using the
definition of $\heta$ and the continuity equation written in Fourier
space, a $2 \times 2$ algebraic system can be written and solved
analytically for the unknowns $\hu$ and $\hw$. Its solution is
available in analytical form only when the variables are
Fourier-transformed in the homogeneous directions. The present method
therefore enjoys its computational efficiency only when a Fourier
discretization is employed for these directions, which means that
either periodic boundary conditions are suitable for the physical
problem under consideration or a fringe-region technique
\cite{bertolotti-herbert-spalart-1992} is adopted.

The unknowns are represented in terms of truncated Fourier series in
the homogeneous directions.  For example the wall-normal velocity
component $v$ is represented as:
\begin{equation}
\label{eq:cart-discretization}
v(x,z,y,t)=\sum_{h=-nx/2}^{+nx/2} \; \sum_{\ell=-nz/2}^{+nz/2}
\hv_{h \ell}(y,t) \rme^{i h \alpha_0 x} \rme^{i \ell \beta_0 z}
\end{equation}
where $h$ and $\ell$ are integer indices corresponding to the
streamwise and spanwise direction respectively, and $\alpha_0$ and
$\beta_0$ are the corresponding fundamental wavenumbers, defining the
streamwise and spanwise periods $L_x = 2 \pi /
\alpha_0 $ and $L_z = 2 \pi / \beta_0 $ of the computational domain.  

The numerical evaluation of the velocity products would require
computationally expensive convolutions in wavenumber space, but can be
carried out efficiently by transforming the three Fourier components
of velocity back into physical space, multiplying them in all six
possible pair combinations, and eventually re-transforming the six
results into wavenumber space. Fast-Fourier-Transform (FFT) algorithms
are used in both directions. This technique is often considered
``pseudo-spectral'', but it should be observed that, when de-aliasing
is performed by expanding the number of collocation points by a factor
of at least 3/2 before going from wavenumber space into physical
space, the velocity products become exactly identical to the
``spectral'' ones that could have been obtained, at a much higher
computational cost, through the actual evaluation of the convolution
products.

\subsection{Time discretization}
\label{sec:timescheme}

Time integration of the equations is performed by a partially-implicit
method, implemented in such a way as to reduce the memory requirements
of the code to a minimum, by exploiting the finite-difference
discretization of the wall-normal direction. The use of a
partially-implicit scheme is a common approach in DNS
\cite{kim-moin-moser-1987}: the explicit part of the equations can
benefit from a higher-accuracy scheme, while the stability-limiting
viscous part is subjected to an implicit time advancement, thus
relieving the stability constraint on the time-step size $\Delta
t$. We employ an explicit third-order, low-storage Runge--Kutta
method, combined with an implicit second-order Crank-Nicolson scheme
\cite{moser-kim-mansour-1999,kim-2003}. 

The procedure to solve the discrete equations for $\hv_{h,\ell}^{n+1}$
and $\heta_{h,\ell}^{n+1}$ at the time level $n+1$ is made by two
distinct steps. In the first step, the RHSs corresponding to the
explicit part have to be assembled. In the representation
(\ref{eq:cart-discretization}), at a given time the Fourier
coefficients of the variables are represented at different $y$
positions; hence the velocity products can be computed through
inverse/direct FFT in wall-parallel planes. Their spatial derivatives
are then computed: spectral accuracy can be achieved for wall-parallel
derivatives, whereas the finite-differences compact schemes described
in \S\ref{sec:cart-fd} are used in the wall-normal direction.  These
spatial derivatives are eventually combined with values of the RHS at
previous time levels. The whole $y$ range from one wall to the other
must be considered.

The second step involves, for each $\alpha, \beta$ pair, the solution
of a set of two ODEs, derived from the implicitly integrated viscous
terms, for which the RHS is now known.  A finite-differences
discretization of the wall-normal differential operators produces two
real banded matrices, in particular pentadiagonal matrices when a
5-point stencil is used. The solution of the resulting two linear
systems gives $\heta_{h \ell}^{n+1}$ and $\hv_{h
\ell}^{n+1}$, and then the planar velocity components $\hu_{h
\ell}^{n+1}$ and $\hw_{h \ell}^{n+1}$ can be computed. 
For each $\alpha, \beta$ pair, the solution of the two ODEs requires
the simultaneous knowledge of the RHS in all $y$ positions. The whole
$\alpha,
\beta$ space must be considered. In the $\alpha-\beta-y$ space the 
first step of this procedure proceeds per wall-parallel planes, while
the second one proceeds per wall-normal lines.

\begin{figure}
\centering
\psfrag{j}{$j=1..ny-1$}
\psfrag{FFT}{IFT/FFT}
\psfrag{N(f)}{$\vect{nl} = \mathcal{N}(\vect{f})$}
\psfrag{t}{$t=t+\Delta t$}
\psfrag{rhs-old}{$rhs = \alpha \vect{f} 
+ \beta \vect{nl} + \gamma A \cdot \vect{f}$}
\psfrag{f-old}{$\vect{f} = \theta \ rhs + \xi \ \vect{rhsold}$}
\psfrag{swap-old}{$\vect{rhsold}=rhs$}
\psfrag{lin-old}{solve $(A + \lambda I) \vect{f}=\vect{f}$}
\includegraphics[width=0.65\textwidth]{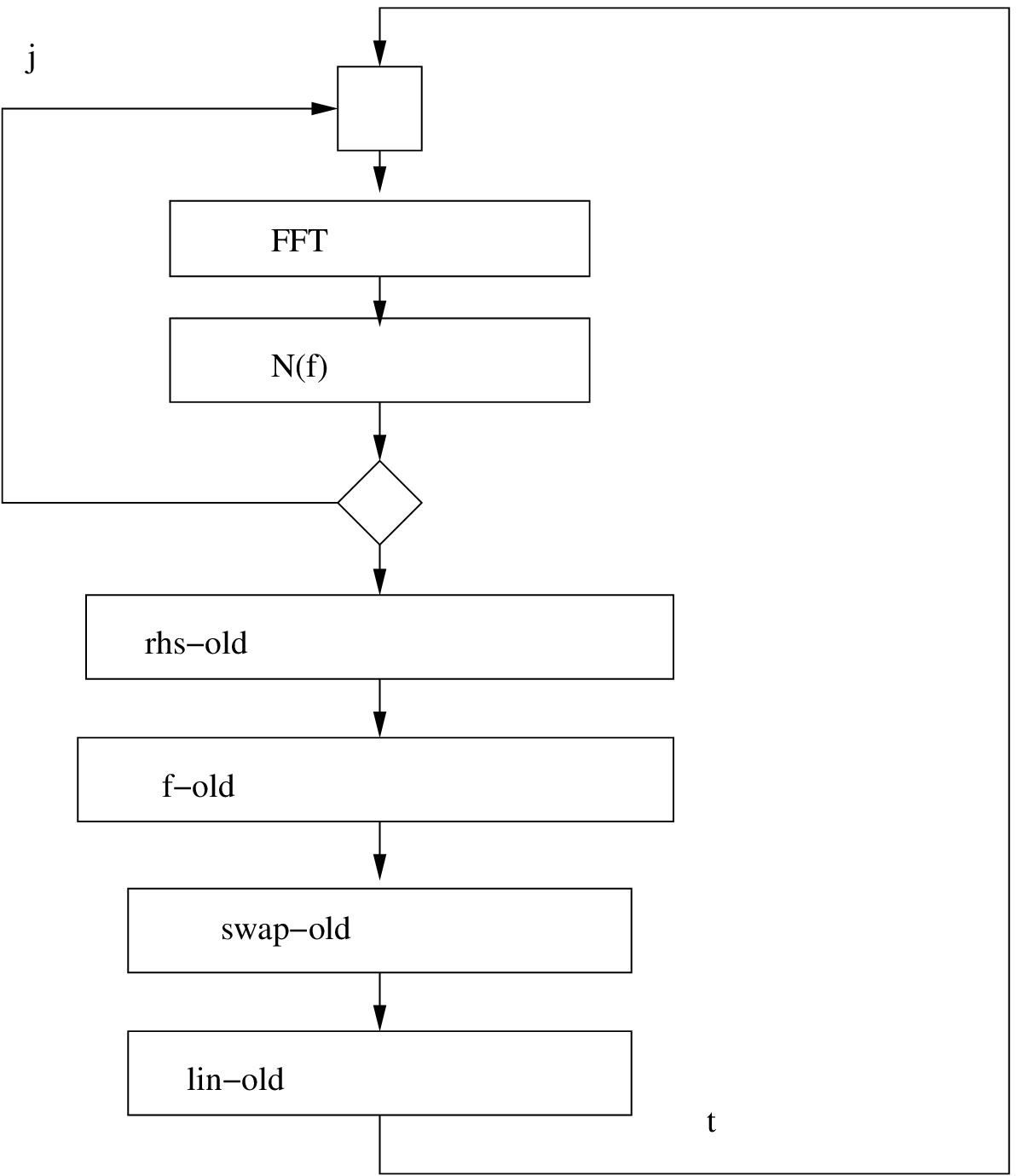}\\
\vspace{1cm}
\psfrag{rhs-new}{$rhs = \alpha \vect{f} 
+ \beta \mathcal{N}(\vect{f})
+ \gamma A \cdot \vect{f}$}
\psfrag{f-new}{$\vect{f} = \theta \ rhs + \xi \ \vect{rhsold}$}
\psfrag{swap-new}{$\vect{rhsold}=rhs$}
\psfrag{lin-new}{solve $(A + \lambda I) \vect{f}=\vect{f}$}
\includegraphics[width=0.65\textwidth]{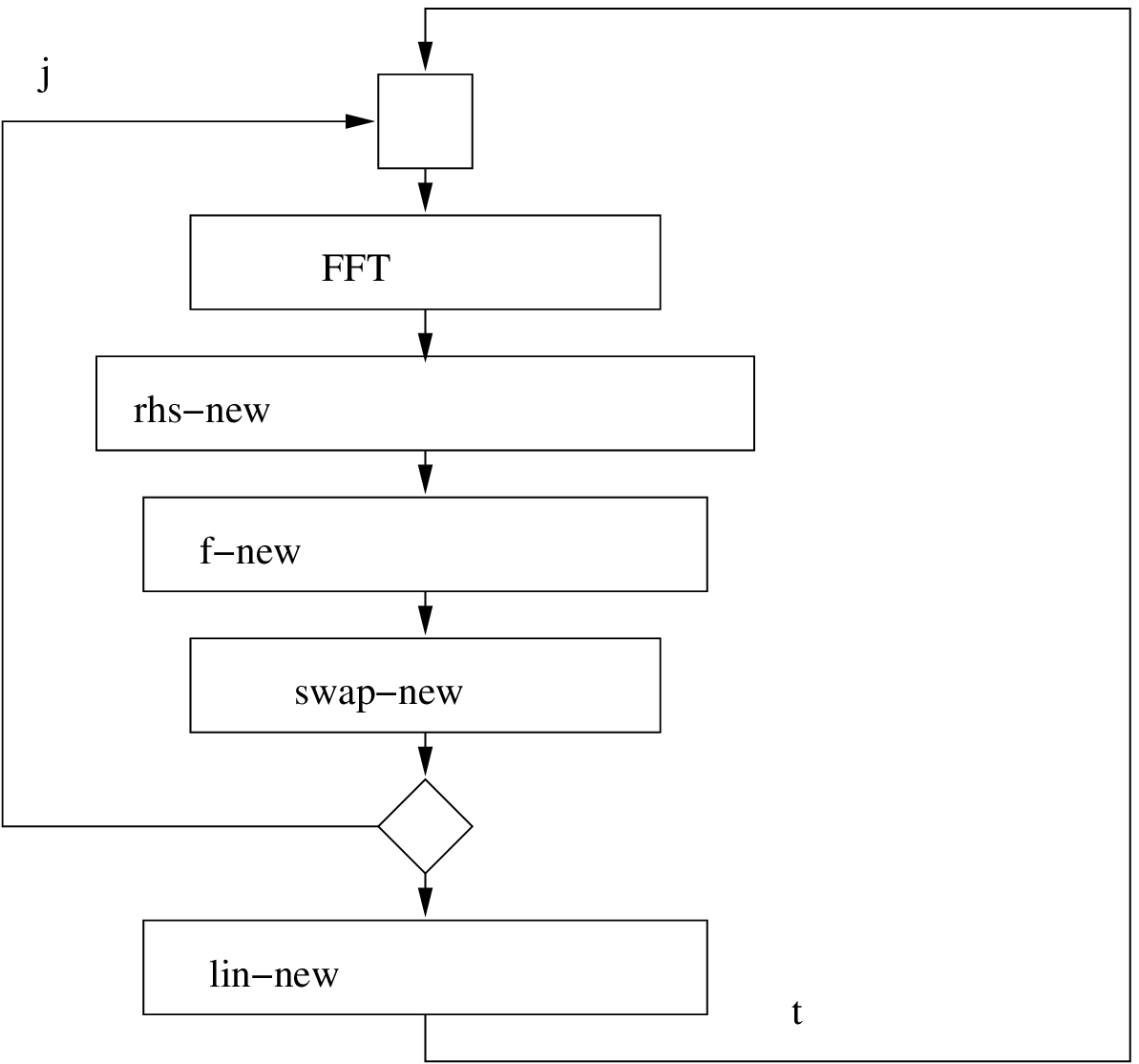}
\caption{Comparison between the standard implementation of a
two-level time-advancement scheme (top), and the present,
memory-efficient implementation (bottom).  Variables printed in bold
require three-dimensional storage space, while italics marks temporary
variables which can use two-dimensional arrays. Greek letters denote
coefficients defining a particular time scheme. The present
implementation reduces the required memory space for a single equation
from 3 to 2 three-dimensional variables.}
\label{fig:timescheme}
\end{figure}

To understand our memory-efficient implementation of the time
integration procedure, let us consider the following differential
equation for the one-dimensional vector $\vect{f}=\vect{f}(y)$:
\begin{equation}
\label{eq:example-timescheme}
\frac{d \vect{f}}{dt} = \mathcal{N}(\vect{f}) + A \cdot \vect{f} ,
\end{equation}
where $\mathcal{N}$ denotes non-linear operations on $\vect{f}$, and
$A$ is the coefficient matrix which describes the linear part. After
time discretization of this generic equation, that has identical
structure to both the $\heta$ and $\hv$ equations, the unknown at time
level $n+1$ stems from the solution of the linear system:
\begin{equation}
\label{eq:example-system}
(A + \lambda I) \cdot \vect{f} = \vect{g}
\end{equation}
where $\vect{g}$ is given by a linear combination (with suitable
coefficients which depend on the particular time integration scheme
and, in the case of Runge-Kutta methods, on the particular sub-step
too) of $\vect{f}$, $\mathcal{N}(\vect{f})$ and $A \cdot \vect{f}$
evaluated at time level $n$ and at a number of previous time
levels. The number of previous time levels depends on the chosen
explicit scheme. For the present, low-storage Runge-Kutta scheme, only
the additional level $n-1$ is required.

The quantities $\vect{f}$, $\mathcal{N}(\vect{f})$ and $A \cdot
\vect{f}$ can be  stored in distinct  arrays, thus resulting in a 
memory requirement of 7 variables per point for a two-levels time
integration scheme. An obvious, generally adopted optimization is the
incremental build into the same array of the linear combination of
$\vect{f}$, $\mathcal{N}(\vect{f})$ and $A \cdot \vect{f}$, as soon as
the single addendum becomes available. The RHS can then be efficiently
stored in the array $\vect{f}$ directly, thus reducing the memory
requirements down to 3 variables per point.

The additional optimization we are able to enforce here relies on the
finite-difference discretization of the wall-normal
derivatives. Referring to our simple example, the incremental build of
the linear combination is performed contemporary to the computation of
$\mathcal{N}(\vect{f})$ and $A \cdot \vect{f}$, the result being
stored into the same array which already contained $\vect{f}$. The
finite-difference discretization ensures that, when dealing with a
given $y$ level, only a little slice of values of $\vect{f}$, centered
at the same $y$ level, is needed to compute
$\mathcal{N}(\vect{f})$. Hence just a small additional memory space,
of the same size of the finite-difference stencil, must be provided,
and the global storage space reduces to two variables per point for
the example equation (\ref{eq:example-timescheme}).

The structure of the time integration procedure implemented in our DNS
code is symbolically shown in the bottom chart of figure
\ref{fig:timescheme}, and compared with the
standard approach, illustrated in the top chart. Within the latter
approach, in a main loop over the wall-parallel planes (integer index
$j$) the velocity products are computed pseudo-spectrally with planar
FFT, their spatial derivatives are taken and the result is eventually
stored in the three-dimensional array $\vect{nl}$. After the loop has
completed, the linear combination of $\vect{f}$, $\vect{nl}$ and $A
\cdot \vect{f}$ is assembled in a temporary two-dimensional array
$rhs$, then combined into the three-dimensional array $\vect{f}$ with
the contribution from the previous time step, and eventually stored in
the three-dimensional array $\vect{rhsold}$ for later use. The RHS,
which uses the storage space of the unknown itself, permits now to
solve the linear system which yields the unknown at the future time
step, and the procedure is over, requiring storage space for 3
three-dimensional arrays.

The flow chart on the bottom of figure
\ref{fig:timescheme} illustrates the present approach. In the main
loop over wall-parallel planes, not only the non-linear terms are
computed, but the RHS of the linear system is assembled plane-by-plane
and stored directly in the three-dimensional array $\vect{f}$,
provided the value of the unknown in a small number of planes (5 when
a 5-point finite-difference stencil is employed) is conserved. As a
whole, this procedure requires only 2 three-dimensional arrays for
each scalar equation.

\subsection{High-accuracy compact finite difference schemes}
\label{sec:cart-fd}

The discretization of the first, second and fourth wall-normal
derivatives $D_1$, $D_2$ and $D_4$, required for the numerical
solution of the present problem is performed through
finite-differences (FD) compact schemes
\cite{lele-1992}. One important difference with \cite{lele-1992} is that our compact schemes are at the same time explicit and at fourth-order accuracy. The computational
molecule is composed of five arbitrarily spaced (with smooth
stretching) grid points on a mesh of $ny+1$ points $y_j$, with $0 \le
j \le ny$. We indicate here with $d^j_1(i), ~ i=-2,\ldots,2$ the five
coefficients discretizing the exact operator $D_1$ over five adjacent
grid points centered at $y_j$, i.e.:
\[
\left. D_1(f(y)) \right|_{y_j} \simeq \sum_{i=-2}^2 d^j_1(i) f(y_{j+i})
\]
where $y_j$ is the $y$ position on the computational mesh where the
derivative has to be evaluated. The coefficients $d^j$ change with the
distance from the wall (i.e. with the integer index $j$) when a
non-uniform mesh is employed.

Compact schemes are also known as implicit finite-differences schemes,
because they typically require the inversion of a linear system for
the actual calculation of a derivative
\cite{lele-1992,mahesh-1998}: this  increases  the complexity and the 
computational cost of such an approach. For the present problem we are
able however to determine explicitly the coefficients for compact,
fourth-order accurate schemes, thanks to the absence of the $D_3$
operator from the present equations. This important simplification has
been highlighted first in the original Gauss-Jackson-Noumerov compact
formulation exploited in his seminal work by Thomas
\cite{thomas-1953}, concerning the numerical solution of the
Orr-Sommerfeld equation.

To illustrate Thomas' method, let us consider a 4th-order ordinary
differential equation (linear for simplicity) for a function $f(y)$ in
the following conservation form:
\begin{equation}
\label{eq:example-eq}
D_4 \left( a_4 f \right) + D_2 \left( a_2 f \right) +
D_1 \left( a_1 f \right) + a_0 f = g
\end{equation}
where the coefficients $a_i=a_i(y)$ are arbitrary functions of the
independent variable $y$, and $g=g(y)$ is the known RHS. Let us
moreover suppose that in frequency space a differential operator, for
example $D_4$, is approximated as the ratio of two polynomials, say
$\sD_4$ and $\sD_0$.  Polynomials like $\sD_4$ and $\sD_0$ have their
counterpart in physical space, and $d_4$ and $d_0$ are the
corresponding FD operators.  The key point is to impose that {\em all}
the differential operators appearing in the example equation
(\ref{eq:example-eq}) admit a representation such as the preceding
one, in which the polynomial $\sD_0$ at the denominator remains {\em
the same}. Eq. (\ref{eq:example-eq}) can thus be recast in the new,
equivalent discretized form:
\begin{equation}
\label{eq:d0}
d_4 \left( a_4 f \right) + d_2 \left( a_2 f \right) +
d_1 \left( a_1 f \right) + d_0 \left( a_0 f \right) = d_0 \left( g
\right)
\end{equation}
and this allows us to use explicit FD schemes, provided the operator
$d_0$ is applied to the RHS of the equation and to the terms not
involving $y$ derivatives. The overhead related to the use of implicit
finite difference schemes disappears, while the advantage of using
compact schemes is retained.

When compared to \cite{thomas-1953}, the present approach is similar,
but we decided to allow for variable coefficients $a_i(y)$ inside the
differential operators $D_i$. Thomas' choice of considering
differential operators of the form $a_i D_i(f)$ is equivalent in
principle, but it would have required to solve for an auxiliary
variable $f'=d_0(f)$. The present choice moreover is better suited for
a differential equation written in conservative form, where only 6
convolutions have to be evaluated.

In our implementation, to obtain a formal accuracy of order 4 we have
used a computational stencil of five grid points. To compute the
finite-difference coefficients, we have followed a standard procedure
in the theory of Pad\'e approximants
\cite{pozzi-1994}. For each distance
$y_j$ from the wall, a $10 \times 10$ linear system can be set up and
solved for the unknown coefficients. A mesh with variable size in the
wall-normal direction is often desirable, in order to keep track of
the increasingly smaller turbulence length scales when the wall is
approached. The use of a non-uniform mesh together with compact
schemes at high accuracy is known
\cite{spotz-carey-1998} to require special care when  the 
differential equation is used as an additional relation which can be
differentiated to eliminate higher-order truncation errors. In the
present approach the use of a non-uniform mesh in such a way as to
still keep a fourth-order accuracy simply requires the procedure
outlined above to be performed (numerically) again for each $y$
station, but only once at the beginning of the computations. The
computer-based solution of these systems requires a negligible
computing time.

We end up with FD operators which are altogether fourth-order
accurate; the sole operator $D_4$ is discretized at sixth-order
accuracy. As suggested in
\cite{lele-1992} and \cite{mahesh-1998}, the use of all the degrees of
freedom for achieving the highest formal accuracy might not always be
the optimal choice. We have therefore attempted to discretize $D_4$ at
fourth-order accuracy only, and to spend the remaining degree of
freedom to improve the spectral characteristics of {\em all} the FD
operators at the same time. Our search has shown however that no
significant advantage can be achieved: the maximum of the errors can
be reduced only very slightly, and - more important - this reduction
does not carry over to the entire frequency range.

The boundaries obviously require non-standard schemes to be designed
to properly compute derivatives at the wall. For the boundary points
we use non-centered schemes, whose coefficients can be computed
following the same approach as the interior points, thus preserving by
construction the formal accuracy of the method. Nevertheless the
numerical error contributed by the boundary presumably carries a
higher weight than interior points, albeit mitigated by the
non-uniform discretization. A systematic study of this error
contribution and of alternative more refined treatments of the
boundary are ongoing work.

\section{The parallel strategy}
\label{sec:parallel}

\subsection{Distributed-memory computers}
\label{sec:distributed-mem}

If the calculations are to be executed in parallel by $p$ computing
machines (nodes), data necessarily reside on these nodes in a
distributed manner, and communication between nodes will take
place. Our main design goal is to keep the required amount of
communication to a minimum.

When a fully spectral discretization is employed, a transposition of
the whole dataset across the computing nodes is needed every time the
numerical solution is advanced by one time (sub)step when non-linear
terms are evaluated. This is illustrated for example in the paper by
Pelz \cite{pelz-1991}, where parallel FFT algorithms are discussed in
reference to the pseudo-spectral solution of the Navier--Stokes
equations. Pelz shows that there are basically two possibilities,
i.e. using a distributed FFT algorithm or actually transposing the
data, and that they essentially require the same amount of
communication. The two methods are found in \cite{pelz-1991} to
perform, when suitably optimized, in a comparable manner, with the
distributed strategy running in slightly shorter times when a small
number of processors is used, and the transpose-based method yielding
an asymptotically faster behavior for large $p$. The large amount of
communication implies that very fast networking hardware is needed to
achieve good parallel performance, and this restrict DNS to be carried
out on very expensive computers only.

Of course, when a FD discretization in the $y$ direction is chosen
instead of a spectral one, it is conceivable to distribute the
unknowns in wall-parallel slices and to carry out the two-dimensional
inverse/direct FFTs locally to each machine. Moreover, thanks to the
locality of the FD operators, the communication required to compute
wall-normal spatial derivatives of velocity products is fairly small,
since data transfer is needed only at the interface between contiguous
slices.  The reason why this strategy has not been used so far is
simple: a transposition of the dataset seems just to have been delayed
to the second half of the time step advancement procedure. Indeed, the
linear systems which stem from the discretization of the viscous terms
require the inversion of banded matrices, whose principal dimension
span the entire width of the channel, while data are stored in
wall-parallel slices.

\begin{figure}
\centering
\psfrag{alpha}{$\alpha$}
\psfrag{beta}{$\beta$}
\psfrag{y}{$y$}
\includegraphics[width=0.7\columnwidth]{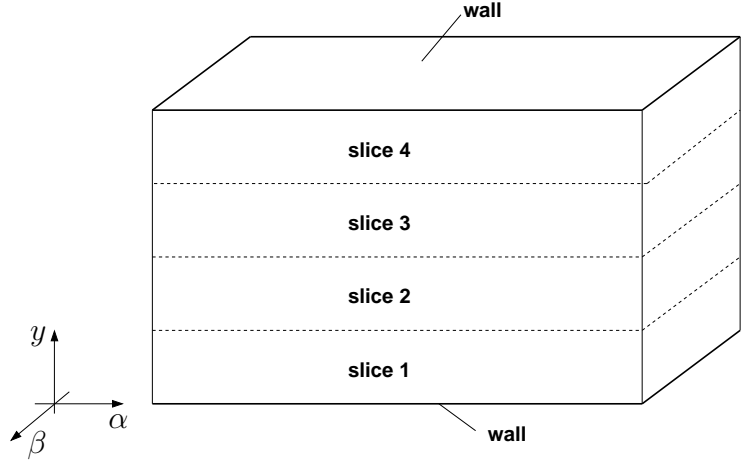}
\caption{Arrangement of data in  wall-parallel slices across the
channel, for a parallel execution with $p=4$ computing nodes.}
\label{fig:slices}
\end{figure}

A transpose of the whole flow field can be avoided however when data
are distributed in slices parallel to the walls, with FD schemes being
used for wall-normal derivatives. The arrangement of the data across
the machines is schematically shown in figure~\ref{fig:slices}: each
machine holds all the streamwise and spanwise wavenumbers for $ny/p$
contiguous $y$ positions. As said, the planar FFTs do not require
communication at all. Wall-normal derivatives needed for the
evaluation of the RHSs do require a small amount of communication at
the interface between contiguous slices.  However, this communication
can be avoided at all if, when using a 5-point stencil, two boundary
planes on each internal slice side are duplicated on the neighboring
slice. This duplication is obviously a waste of computing time, and
translates into an increase of the actual size of the computational
problem. However, since the duplicated planes are $4 (p-1)$, as long
as $p \ll ny$ this overhead is negligible. When $p$ becomes
comparable to $ny$, an alternative procedure involving a small amount
of communication becomes convenient. We will further discuss this
point in \S\ref{sec:performance}.

The critical part of the procedure lies in the second half of the
time-step advancement, i.e. the solution of the set of two linear
systems, one for each $h,\ell$ pair, and the recovery of the planar
velocity components: the necessary data just happen to be spread over
all the $p$ machines. It is relatively easy to avoid a global
transpose, by solving each system in a {\em serial} way across the
machines: adopting a LU decomposition of the pentadiagonal,
distributed matrices, and a subsequent sweep of back-substitutions,
only a few coefficients at the interface between two neighboring nodes
must be transmitted. The global amount of communication remains very
low and, at the same time, local between nearest neighbors only. The
problem here is obtaining a reasonably high parallel efficiency: if a
single system had to be solved, the computing machines would waste
most of their time waiting for the others to complete their task. In
other words, with the optimistic assumption of infinite communication
speed, the total wall-clock time would be simply equal to the
single-processor computing time.

The key observation to obtain high parallel performance is that the
number of linear systems to be solved at each time (sub)step is very
large, i.e.  $(nx+1) (nz + 1)$, which is at least $10^4$ and
sometimes much larger in typical DNS calculations
\cite{delalamo-jimenez-2003}. This allows the solution of the 
linear systems to be efficiently pipelined as follows. When the LU
decomposition of the matrix of the system for a given $h, \ell$ pair
is performed (with a standard Thomas algorithm adapted to
pentadiagonal matrices), there is a first loop from the top row of the
matrix down to the bottom row (elimination of the unknowns), and then
a second loop in the opposite direction (back-substitution). The
machine owning the first slice performs the elimination in the local
part of the matrix, and then passes on the boundary coefficients to
the neighboring machine, which starts its elimination.  Instead of
waiting for the elimination in the $h, \ell$ system matrices to be
completed across the machines, the first machine can now immediately
start working on the elimination in the matrix of the following
system, say $h, \ell+1$, and so on. After the elimination in the first
$p$ systems is started, all the computing machines work at full
speed. A synchronization is needed only at the end of the elimination
phase, and then the whole procedure can be repeated for the
back-substitution phase.

Clearly this pipelined-linear-system (PLS) strategy involves an
inter-node communication made by frequent sends and receives of small
data packets (typically two lines of a pentadiagonal matrix, or two
elements of the RHS array).  While the global amount of data is very
low, this poses a serious challenge to out-of-the-box communication
libraries, which are known to have a significant overhead for very
small data packets. In fact, as we will mention in
\S\ref{sec:performance}, we have found unacceptably poor performance
when using MPI-type libraries. On the other hand we have succeeded in
developing an effective implementation of inter-node communication
using only the standard i/o functions provided by the C
library. Details of this alternative implementation are illustrated in
\S\ref{sec:personal-supercomputer}.

\subsection{Shared-memory machines}
\label{sec:shared-mem}

The single computing node may be single-CPU or multi-CPU. In the
latter case, it is possible to exploit an additional and complementary
parallel strategy, which does not rely on message-passing
communication anymore, and takes advantage of the fact that local CPUs
have direct access to the same, local memory space.  We stress that
this is different from using a message-passing strategy on a
shared-memory machine, where the shared memory simply becomes a faster
transmission medium. Using multiple CPUs on the same memory space may
yield an additional gain in computing time, at the only cost of having
the computing nodes equipped with more than one (typically two)
CPUs. For example the FFT of a whole plane from physical to
Fourier-space and vice-versa can be easily parallelized this way, as
well as the computing-intensive part of building up the RHS terms.
With SMP machines, high parallel efficiencies can be obtained quite
easily by ``forking'' new processes which read from and write to the
same memory space; the operating system itself then handles the
assignment of tasks to different CPUs, and only task synchronization
is a concern at the programming level.

\subsection{The Personal Supercomputer}
\label{sec:personal-supercomputer}

\begin{figure}
\centering
\includegraphics[width=0.6\columnwidth]{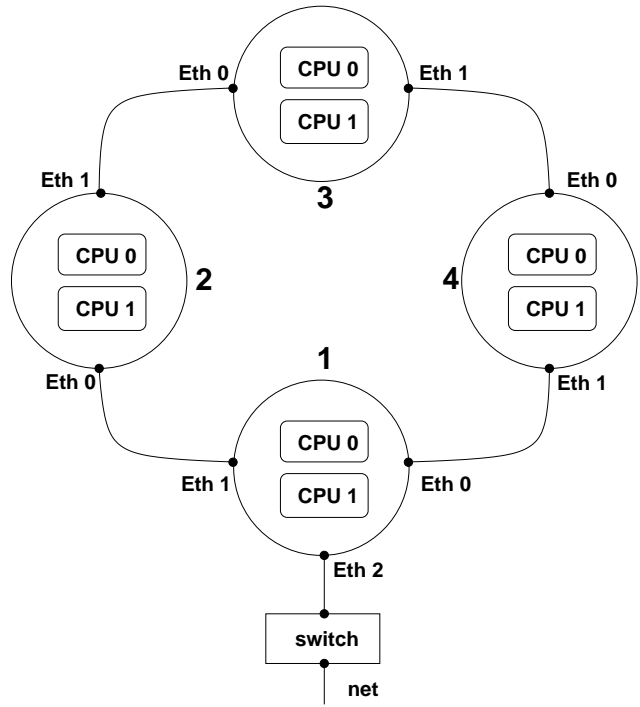}
\caption{Conceptual scheme of the connection topology for  a 
computing system made by 4 nodes; one machine may be connected to the
local net through a switch, if the system has to be operated
remotely.}
\label{fig:cluster}
\end{figure}

While a computer program based on the numerical method described
heretoforth can be easily used on a general-purpose cluster of
machines, connected through a network and a switch, for maximum
efficiency a dedicated computing system can be specifically designed
and built on top of the parallel algorithm described above.

At the CPU level, the mass-marketed CPUs which are commonly found
today in desktop systems are the perfect choice: their performance is
comparable to the computing power of the single computing element of
any supercomputer \cite{dongarra-2004}, at a fraction of the
price. The single computing node can hence be a standard desktop
computer; SMP mainboards with two CPUs are very cheap and easily
available.

The present PLS parallel strategy allows an important simplification
in the connection topology of the machines. Since the transposition of
the whole dataset is avoided, communications are always of the {\em
point-to-point} type; moreover, each computing machine needs to
exchange data with and only with two neighboring machines only. This
can be exploited with a simple ring-like connection topology among the
computing machines, sketched in figure~\ref{fig:cluster}, which
replicates the logical exchange of information and the data structure
previously illustrated in figure~\ref{fig:slices}: each machine is
connected through two network cards only to the previous machine and
to the next. The necessity of a switch (with the implied additional
latency in the network path) is thus eliminated, in favor of
simplicity, performance and cost-effectiveness.

Concerning the transmission protocol, the simplest choice is the
standard, error-corrected TCP/IP protocol. We have estimated that on
typical problem sizes the overall benefits from using a dedicated
protocol (for example the GAMMA protocol described in
\cite{ciaccio-chiola-1999}) would be negligible: since the ratio
between communication time and computing time is very low (see
\S\ref{sec:performances-parallel} and Fig.\ref{fig:tempi-itanium}), 
the improvements by using such a protocol are almost negligible, and
to be weighed against the increase in complexity and decrease in
portability.

The simplest and fastest strategy we have devised for the
communication type is to rely directly on the standard networking
services of the Unix operating system, i.e.  sockets (after all,
message-passing libraries are socket-based).  At the programming
level, this operation is very simple, since a socket is seen as a
plain file to write into and to read from. Using sockets allows us to
take advantage easily and efficiently of the advanced buffering
techniques incorporated in the management of the input/output streams
by the operating system: after opening the socket once and for all, it
is sufficient to write (read) data to (from) the socket whenever they
are available (needed), and the operating system itself manages
flushing the socket when its associated buffer is full. We have found
however that for best performances the buffer size had to be
empirically adjusted: for Fast Ethernet hardware, the optimum has been
found at the value of 800 bytes, significantly smaller than the usual
value (the Linux operating system defaults at 8192).

We have built a prototype of such a dedicated system, composed of 8
SMP Personal Computers. Each node is equipped with 2 Pentium III
733MHz CPU and 512MB of 133MHz SDRAM. The nodes are connected to each
other by two cheap 100MBits Fast Ethernet cards. We call such a
machine a Personal Supercomputer.  The performance of our numerical
method used on this system will be shown in \S\ref{sec:performance} to
be comparable to that of a supercomputer. Such machines enjoy the
advantages of a simple desktop Personal Computer: low cost and easy
upgrades, unlimited availability even to a single user, low weight,
noise and heat production, small requirements of floor space, etc.
Further details and instructions to build and configure such a machine
can be found in \cite{quadrio-luchini-2004}.

\section{Performance}
\label{sec:performance}

The performance of the present pipelined-linear-system (PLS) method is
assessed here in terms of memory requirements, single-processor CPU
time and parallel speedup, both in absolute terms and by comparison
with similar DNS codes. We compare it also with an alternative version
of the code, that we have written with a different distributed-memory
parallel strategy. This alternative code employs the more traditional
transpose FFT method \cite{pelz-1991}, so that the streamwise
wavenumbers are distributed across the system while the wall-normal
direction is local to each processor. Both parallel algorithms have
been preliminarly tested for correctness, and checked to give
identical output between a single-processor run and a truly parallel
execution.

Our tests have been mainly conducted on the computing system described
in \S\ref{sec:personal-supercomputer}, either on a single Pentium III
733 MHz CPU for the single-processor tests, or by using multiple
processors for the parallel tests. A few measurements are collected by
using more recent dual-processor Opteron machines, available at
Salerno University. Each of the Opteron machines is equipped with two
1.6 GHz AMD CPUs, and carries 1 GByte of RAM; the machines are
equipped with 3 Gigabit Ethernet cards each. They are thus connected
both in a switched network via one card and the interposed HP 2724
switch, and in the ring topology by means of the other two cards.

The performance of our codes on these machines cannot easily be
compared with performance figures of similar codes, since the
information available in the literature is often partial and based on
each time different computing machines. Some data (for example, RAM
requirements) of course can be compared directly, given their
independence of the particular computer architecture. For other data
(typically, CPU time) we report our own figures, and we try in
addition to compare qualitatively such data with CPU time on different
architectures, by using the Performance Database maintained and
published monthly by Dongarra \cite{dongarra-2004}. 

The results presented in what follows are computed with the parallel
algorithm described in \S\ref{sec:distributed-mem}. As already pointed
out, this algorithm is especially well suited when a limited number of
computing nodes is available. Indeed, the duplication of four
computing planes for each internal slice interface implies a CPU
penalty, while allowing the bare minimum of inter-node
communication. This penalty increases with the number of computing
nodes and decreases with the number of discretization points in the
$y$ direction.

We define the speedup factor as the ratio of the actual wall-clock
computing time $t_p$ obtained with $p$ nodes and the wall-clock time
$t_1$ required by the same computation on a single node:
\[
S(p) = \frac{t_p}{t_1} .
\]

The maximum or ideal speedup factor $S_i$ that we can expect with our
PLS algorithm, corresponding to the assumption of infinite
communication speed, is less than linear, and can be estimated with
the formula:
\begin{equation}
\label{eq:ideal-speedup}
S_i(p) = p \left( 1 - \frac{4(p-1)}{ny} \right),
\end{equation}
where the factor $4$ accounts for the two wall-parallel planes
duplicated at each side of interior
slices. Eq. (\ref{eq:ideal-speedup}) reduces to a linear speedup when
$ny \rightarrow \infty$ for a finite value of $p$. A quantitative
evaluation of the function (\ref{eq:ideal-speedup}) for typical values
of $ny=\mathcal{O}(100)$ shows that the maximum achievable speedup is
nearly linear as long as the number of nodes remains moderate, i.e. $p
< 10$. We are presently considering a slightly different parallel
implementation, still in development at the present time, which is
better suited for use when $p=\mathcal{O}(ny)$.

\subsection{Memory requirements}

One fundamental requirement for a DNS code is to save RAM: Jim\'enez
in \cite{jimenez-2003} considers RAM occupation and CPU time as
performance monitors of equivalent importance. The amount of required
RAM is dictated by the number and the size of the three-dimensional
arrays, and it is typically reported 
\cite{kim-moin-moser-1987,skote-2001,jimenez-2003} to be no less 
than $7 \ nx \times ny \times nz$ floating-point variables. Cases
where RAM requirements are significantly higher are not uncommon: for
example in
\cite{guenther-etal-1998} a channel flow simulation of 
$128 \times 65 \times 128$ reportedly required 1.2GB of RAM,
suggesting a memory occupation approximately 18 times larger.

In our code all the traditional optimizations are employed: for
example there is no reserved storage space for $\heta$, which
overwrites $\hu$ in certain sections of the time-integration
procedure, and is overwritten by $\hu$ in other sections. An
additional saving specific to the present method comes from the
implementation of the time advancement procedure, discussed in
\S\ref{sec:timescheme}, which takes advantage of the finite-difference
discretization of the wall-normal derivatives.  Each of the two scalar
equations for $\heta$ and $\hv$ requires two variables per point. In
addition, solving the algebraic system for $\hu$ and $\hw$ raises the
global memory requirement to 5 variables per point. Thus our code
requires a memory space of $5 \ nx
\times ny \times nz$ floating-point variables, plus  workspace and 
two-dimensional arrays. For example a simulation with $nx = ny = nz =
128$ takes only 94 MBytes of RAM (using 64-bit floating-point
variables).

In a parallel run the memory requirement can be subdivided among the
computing machines. With $p=2$ the same $128^3$ case runs with 53
MBytes of RAM (note that the amount of RAM is slightly larger than one
half of the $p=1$ case, due to the aforementioned duplication of
boundary planes). The system as a whole therefore allows the
simulation of turbulence problems of very large computational size
even with a relatively small amount of RAM deployed in each node. A
problem with computational size of $400^3$ would easily fit into our 8
nodes equipped with 512MB RAM each.

\subsection{CPU requirements}

As far as CPU efficiency is concerned, without special optimization
the $128^3$ test case mentioned above requires 42.8 CPU seconds for
the computation of a full three-sub-steps Runge-Kutta temporal step on
a single Pentium III 733MHz processor. Unfortunately, we are not aware
of papers where a similar code is clearly documented in terms of time
required for running a problem of a specified size on these CPU
types. One can however deduce from \cite{skote-2001} that a
computational case of slightly smaller size, i.e. $128 \times 96
\times 128$ (which takes 31 seconds on our machine) runs on a single
processor of the 256 nodes Cray T3E of the National Supercomputer
Center of Link\"oping (Sweden) in approximately 40 seconds, and on a
single processor of the 152 nodes IBM SP2 machine, available at the
Center for Parallel Computers of KTH University, in 8 seconds of CPU
time. These timings are in a ratio which is not far from the ratio
among the computing power of the different CPUs, as deduced
from the tables reported in
\cite{dongarra-2004}, and indicate that the SP2 machine is the one 
which is able to achieve the higher percentage of its theoretical peak
power. The present code hence is roughly equivalent (in its serial
version) to that described in \cite{skote-2001} in terms of CPU
efficiency. Another paper which reports execution times for a $128^3$
problem is in ref. \cite{iovieno-cavazzoni-tordella-2001}: their code
for isotropic turbulence appears to run on one CPU of an IBM SP3 Power3
Nighthawk taking approximately 10 minutes per time step.

The internal timings of our code show that the direct/inverse
two-dimensional FFT routines take the largest part of the CPU time,
namely 56\%. The calculation of the RHS of the two governing equations
(where wall-normal derivatives are evaluated) takes 25\% of the total
CPU time, the solution of the linear systems arising from the implicit
part around 12\%, and the calculation of the planar velocity
components 3\%. The time-stepping scheme takes 3\% and computing a few
runtime statistics requires an additional 1\% of the CPU time.

\subsection{Parallel efficiency}
\label{sec:performances-parallel}

\subsubsection{Distributed-memory speedup}

\begin{figure}
\center
\includegraphics[width=0.9\textwidth]{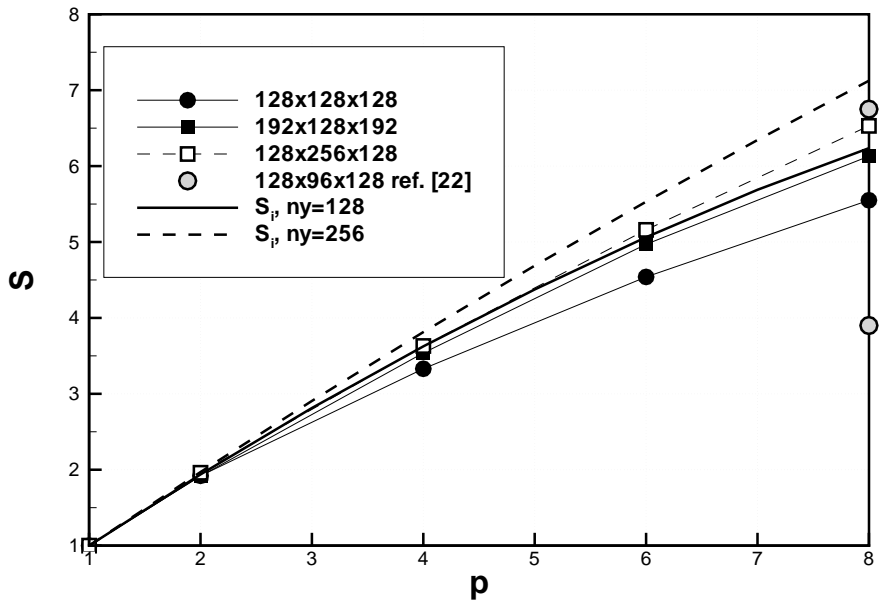}
\caption{Measured speedup on the Pentium III-based machine as 
a function of the number $p$ of computing nodes. Thick lines are the
ideal speedup $S_i$ from Eq.~(\ref{eq:ideal-speedup}) for $ny=128$
(continuous line) and $ny=256$ (dashed line). Gray circles are
speedups inferred from \cite{skote-2001} for a case with $ny=96$ and
measured on two different supercomputers.}
\label{fig:tempi}
\end{figure}

The parallel (distributed-memory) performance of the code is
illustrated in figure~\ref{fig:tempi}, where speedup ratios are
reported as a function of the number of computing nodes. The maximum
possible speedup $S_i$ is shown with thick lines. $S_i$ approaches the
linear speedup for large $ny$, being reasonably high as long as $p$
remains small compared to $ny$: with $p=8$ it is 6.25 for $ny=128$ and
7.125 for $ny=256$.  Notwithstanding the commodity networking hardware
and the overhead implied by the error-corrected TCP protocol, the
actual performance compared to $S_i$ is extremely good, and improves
with the size of the computational problem.  The percentage of time
$t_c$ spent for communication is estimated as follows:
\begin{equation}
\label{eq:percentage-comm-time}
\% \ t_c = 100 \ \frac{t_{p} - t_1/S_i(p)}{t_p}.
\end{equation}

The case $192 \times 128 \times 192$ is hardly penalized by the time
spent for communication, which is only 2\% of the total computing time
when $p=8$. The communication time becomes 7\% of the total computing
time for the larger case of $nx=128$, $ny=256$ and $nz=128$, and is
$12$\% for the worst (i.e. smallest) case of $128^3$, which requires
7.7 seconds for one time step on our machine, with a speedup of
5.55. In figure \ref{fig:tempi} we report also (with gray symbols)
speedup data from \cite{skote-2001} for his case with size $128 \times
96 \times 128$: this case runs in approximately 2 seconds when 8
processors are used on the SP2 (speedup 3.9), while it requires 6
seconds with 8 processors of the T3E (speedup 6.8).

It is worth mentioning again that our communication procedure can easily be
implemented through a standard message-passing library, but shows in this case
a noticeably degraded performance. In fact we have tested the present method
on a $128^3$ case with $p=2$ and using the MPI library. A speedup of $S=0.87$
has been measured  (i.e. the wall clock is increased), to be compared with
$S_i=1.94$ and a measured speedup of $S=1.92$ with the use of plain sockets.
This result is explained by the large overhead implied by the MPI library, that
is known to be inefficient in transmitting extremely small data packets.  We
would like to stress again that, from a programming point of view, plain
sockets are definitely simple to use: once the socket is properly opened, the
procedure of communication with another machine boils down to simply writing to
or reading from a file.

\begin{figure}
\center
\includegraphics[width=0.9\textwidth]{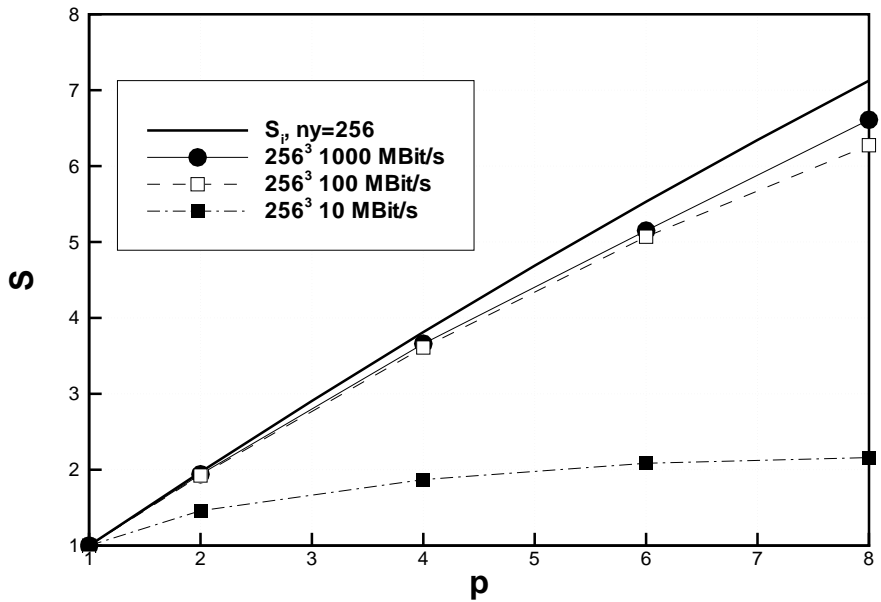}
\caption{Measured speedup on the Opteron-based 
machine as a function of the number $p$ of computing nodes. Thick line
is the ideal speedup from Eq. (\ref{eq:ideal-speedup}) for
$ny=256$. Speedup measured when using Gigabit Ethernet cards
(circles), and the same cards run at the slower speed of 100MBit/s
(empty squares) and 10 MBit/s (filled squares).}
\label{fig:tempi-cfd}
\end{figure}

Figure \ref{fig:tempi-cfd} illustrates the speedup achieved with the
faster Opteron machines connected via Gigabit Ethernet cards in the
ring-topology layout, compared with $S_i$. The test case has a size of
$256^3$. The CPUs of this system are significantly faster than the
Pentium III, and the network cards, while having 10 times larger
bandwidth, have latency characteristics typical of Fast Ethernet
cards. It is remarkable how well the measured speedup still approaches
the ideal speedup, even at the largest tested value of
$p$. Furthermore, we report also the measured speedup when the Opteron
machines are used with the Gigabit cards set up to work at the lower
speeds of 100 MBit/s and 10MBit/s. It is interesting to observe how
slightly performance is degraded in the case at 100MBit/s, whose curve
is nearly indistinguishable form that at 1GBit/s. Even with the
slowest 10MBit/s bandwidth connecting such fast processors, and with a
problem of large computational size, it is noteworthy how the present
method is capable to achieve a reasonable speedup for low $p$ and not
to ever degrade below $S=1$. This relative insensitivity to the
available bandwidth can be ascribed to the limited amount of
communication required by the present method.  

The amount of data which has to be exchanged by each machine for the
advancement of the solution by one time step made by 3 Runge--Kutta
substeps can be quantified as follows. The number of bytes $D_r$
transmitted and received by each computing node for $p>2$ and in one
complete time step is:
\[
D_r = 3 \times 8 \times nx \times nz \times 88 = 2112 \ nx \times nz
\]
where 3 is the number of temporal substeps, 8 accounts for 8-bytes
variables, and 88 is the total number of scalar variables that are
exchanged at the slice interfaces for each wavenumber pair (during
solution of the linear systems and of the algebraic system to compute
$\hu$ and $\hw$). For the $128^3$ case, $D_r \approx 276$ MBit of
network traffic, evenly subdivided between the two network cards.

Interestingly, the quantity $D_r$ is linear in the total number of
Fourier modes $nx \times nz$, and does not depend upon $ny$. Moreover,
the amount of traffic does not increase when $p$ increases. This has
to be contrasted with the amount of communication required by other
parallel methods; this comparison will be discussed in the next
subsection.

\begin{figure}
\centering
\psfrag{Np}{$p$}
\includegraphics[width=0.9\textwidth]{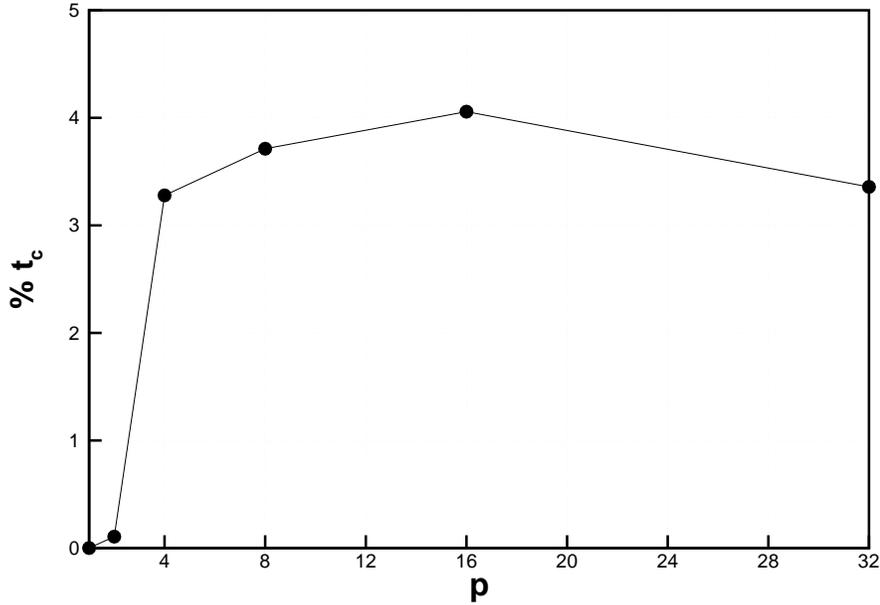}
\caption{Time \%$t_c$ spent for communication as percentage of the 
total computing time, defined in formula
(\ref{eq:percentage-comm-time}), for the Itanium II cluster with
Gigabit Ethernet interconnects. Problem size: $nx=512$, $ny=256$ and
$nz=512$.}
\label{fig:tempi-itanium}
\end{figure}

As already mentioned, the parallel strategy described here targets
only systems where the number $p$ of computing nodes is small compared
to the number $ny$ of points in the $y$ direction. Increasing $p$ at
fixed $ny$ leads to gradually worse performance, since the actual size
of the problem increases owing to the duplicated planes at the
interface. Nevertheless, figure \ref{fig:tempi-itanium} shows that,
even when a larger number of computing nodes is employed on a problem
of a large size, the percentage of the computing time spent for
communication, estimated according to
Eq. (\ref{eq:percentage-comm-time}), remains very low. This figure
reports the parallel performance measured on a cluster of SMP machines
equipped with 2 Itanium II processors, and connected with Gigabit
Ethernet cards and a switch, available courtesy of the SHARCNET
Computing Centre at the University of Western Ontario. The test case
has a size of $nx=512$, $ny=256$ and $nz=512$.

Figure~\ref{fig:tempi-itanium} suggests that the communication
time remains limited up to relatively large values of $p$, even when
the networking hardware (Gigabit Ethernet) allows a bandwidth more
than two orders of magnitude smaller than a typical supercomputer.

\subsubsection{Comparison with a different parallel strategy: 
the transpose method}
\label{sec:transpose}

In order to be able to compare the present PLS parallel strategy with
a standard strategy which performs a block transpose before and after
each FFT, we have written a second version of our code, that adopts
the same FD discretization for the wall-normal direction, but
distributes the streamwise wavenumbers across the nodes, i.e.
organizes data in slices parallel to the $\alpha$-axis of figure
\ref{fig:slices}. The serial performance of the two computer codes is
identical, since for $p=1$ they perform the same operations.  Even
though we have obviously put less optimization effort into the
transpose code, written {\em ad hoc} for this test, compared the the
PLS code, we have to mention that the transpose code has been written
with the basic optimizations in mind. In particular, we have taken
care that communications are scheduled in such a way that the machines
are always busy communicating, which is an essential requirement to
achieve high performance with the transpose FFT.

The amount of data (in bytes) $D_t$ which has to be exchanged by each
machine for the complete advancement by one time step with the
transpose-based method is as follows:
\[
D_t = 3 \times 8 \times (p-1) \frac{nx}{p}
\times \frac{3}{2} \frac{nz}{p} \times ny \times 18 = 
648 \ \frac{p-1}{p^2} \ nx \times nz \times ny
\]

Again, the factors 3 and 8 account for the number of temporal substeps
and the 8-bytes variables respectively. In the whole process of
computing non-linear terms 9 scalars have to be sent and received (3
velocity components before IFT and 6 velocity products after FFT); for
each wall-parallel plane, each machine must exchange with each of the
others $p-1$ nodes an amount of $nx \times nz / p^2$ grid cells, and
the factor $3/2$ corresponds to dealiasing in one horizontal direction
(the $3/2$ expansion, and the subsequent removal of higher-wavenumber
modes, in the other horizontal direction can be performed after
transmission).

The ratio between the communication required by the transpose-based
method and the PLS method can thus be written as:
\[
\frac{D_t}{D_r} = 0.307 \ \frac{p-1}{p^2} \ ny
\]
which corresponds to the intuitive idea that the transpose method
exchanges all the variables it stores locally, whereas the PLS method
only exchanges a (small) number of wall-parallel planes, independent
on $ny$ and $p$. Moreover the ratio $D_t/D_r$, being proportional to
$ny$ for a given $p$, is expected to increase with the Reynolds number
of the simulation, since so does the number of points needed to
discretize the wall-normal direction, thus indicating an increase of
PLS efficiency relative to the transpose strategy. More important,
when the transpose-based method is employed, the global amount of
communication that has to be managed by the switch increases with the
number of computing machines and is all-to-all rather than between
nearest neighbors only, so that its performance is expected to degrade
when a large $p$ is used.

For a case sized $128^3$ and $p=2$, $D_t$ amounts to $\approx$ 2700
MBit. This gives a lower bound for the communication time with the
transpose-based method of 27 seconds with Fast Ethernet, to be
compared with a single-processor execution time of $t_1=42.8$ seconds
on a single Pentium III.  Indeed, when tested on the Fast Ethernet
Pentiums, the transpose-based method has not been found to yield any
reduction of the wall-clock computing time: we have measured
$t_2=51.2$ seconds.  This is in line with the previous estimate: $t_2$
stems from a computing time $t_1/2$ plus a communication time of 29.8
seconds, which is near to the time estimate on the basis of $D_t$ and
assuming network cards running at full speed.
\begin{figure}
\centering
\includegraphics[width=0.9\textwidth]{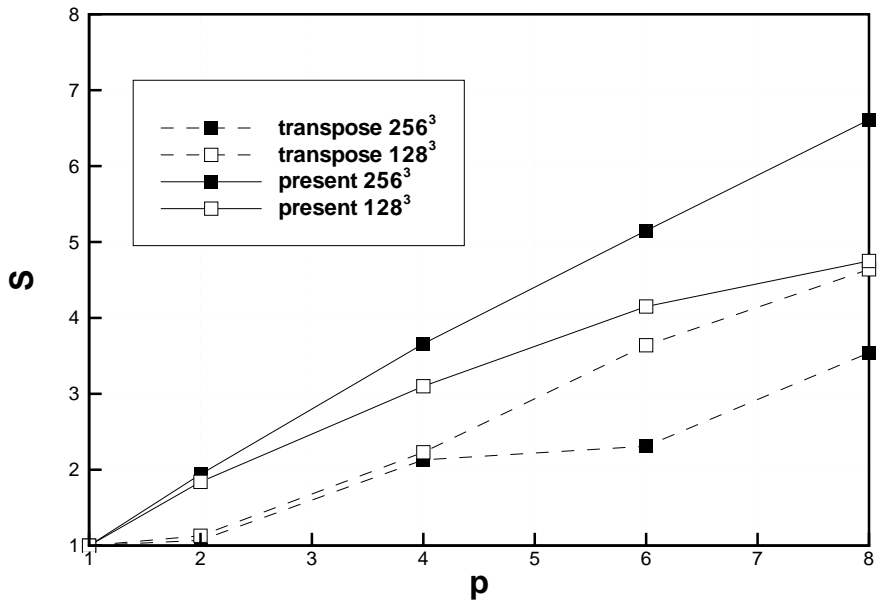}
\caption{Measured speedup on the Opteron-based machine as a function 
of the number $p$ of computing nodes. Continuous line is the PLS
method, and dashed line is the transpose-based method.}
\label{fig:tempi-transpose}
\end{figure}
With a faster network the performance of the transpose-based method
improve. Figure~\ref{fig:tempi-transpose} reports comparative
measurements between the PLS and the transpose-based methods, run on 8
Opteron boxes interconnected with Gigabit Ethernet. The PLS method is
run with the machines connected in a ring, while the transpose-based
method is tested with machines linked through a switch. Measurements
show that $S >1$ can now be achieved with the transpose-based
method. However, the transpose method performs best for the smallest
problem size, while the PLS shows the opposite behavior. For the
$256^3$ case, which is a reasonable size for such machines, the
speedup from the transpose-based method is around one half of what can
be obtained with PLS.

\subsubsection{Shared-memory speedup}

In the present approach, we exploit the availability of two CPUs for
each computing node by assigning to each of them the computation of a
different FFT, and then the RHS setup for one half of the wavenumber
pairs. Since this part of the code is local to each machine, there is
no communication overhead associated with the use of the second CPU,
therefore the gain is essentially independent of the number $p$ of
computing machines.

We have tested machines equipped with 2 CPUs, and measured a 1.55
speedup on a 2-CPU Pentium III box. On the dual Opteron system, which
exploits a faster memory at 400 MHz and a proprietary memory and
inter-processor bus, the speedup increases to 1.7. The implementation
of the shared-memory parallelism was not the main focus of the present
work, so that for simplicity we have parallelized only the non-linear
part of the time-stepping procedure, a portion of the serial code
estimated around 80\%. The additional SMP gain is thus an interesting
result, since it comes at a low cost.  Indeed, the cost of a dual-CPU
node is only a fraction greater than the cost of a single-CPU
computing node. Systems with more than 2 CPU are today significantly
more expensive and, although we have not tested any of them, may be
suspected to perform quite inefficiently when used with the present
application, owing to the increased memory-contention problems. Hence
we regard the use of a second CPU as an added bonus at a little extra
cost.

\begin{figure}
\centering
\includegraphics[width=0.9\textwidth]{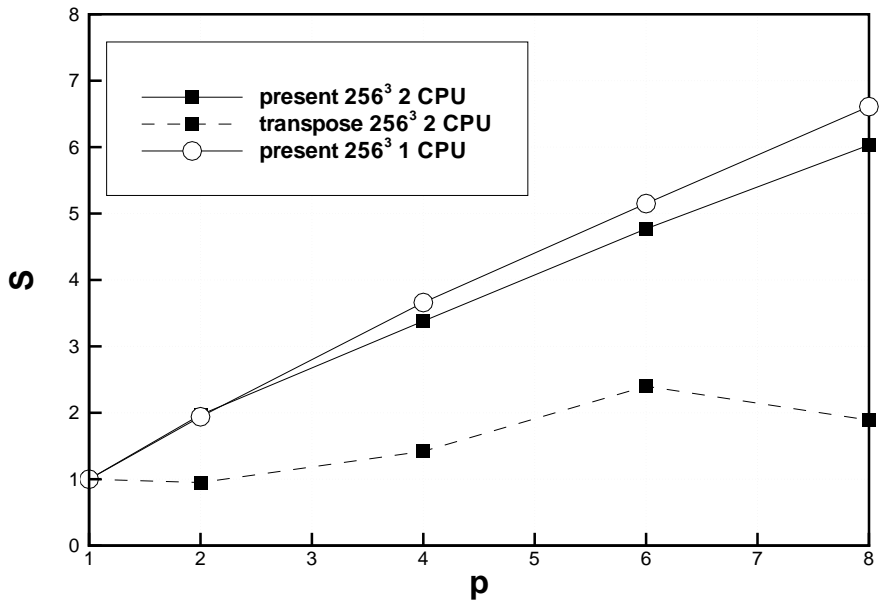}
\caption{Measured speedup on the Opteron-based machine as a function 
of the number $p$ of computing nodes, when 2 CPUs per node are used. A
continuous line denotes the PLS method, and a dashed line the
transpose-based method. Open symbols refer to the single-CPU speedup.}
\label{fig:tempi-transpose-smp}
\end{figure}

The SMP speedup is where the PLS method shows another advantage
compared with the transpose-based method. The second CPU can be
exploited in the transpose-based method too, however in this case the
global effectiveness degrades, since the speed of each node increases
while the communication time remains the same.  This can be
appreciated in figure \ref{fig:tempi-transpose-smp}, where performance
for a case $256^3$ is observed to become unacceptable when $p > 6 $
and two CPU for each machine are used with the transpose method. With
PLS, on the other hand, the penalty associated with the larger amount
of communication per unit of computing time is clearly seen to remain
limited, and no significant decrease of the parallel efficiency is
observed when the second CPU in each node is activated.

\section{Conclusions}
\label{sec:conclusions}

In this paper we have described a numerical method suitable for the
parallel direct numerical simulation of incompressible wall
turbulence, capable of achieving high efficiency by using commodity
hardware. The method can be used when the governing
equations are written either in cartesian or in cylindrical
coordinates.

The key point in its design is the choice of compact finite
differences of fourth-order accuracy for the discretization of the
wall-normal direction. The use of finite differences schemes, while
retaining a large part of the accuracy enjoyed by spectral schemes, is
crucial to the development of the parallel strategy, which exploits
the locality of the FD operators to largely reduce the amount of
inter-node communication. Finite differences are also key to the
implementation of a memory-efficient time integration procedure, which
permits a minimal storage space of 5 variables per point, compared to
the commonly reported minimum of 7 variables per point. This
significant saving is available in the present case too, the use of
compact schemes notwithstanding, since they can be written in explicit
form, leveraging the missing third derivative in the governing
equations.

The parallel method described in this paper, based on the pipelined
solution of the linear systems arising from the discretization of the
viscous terms, achieves its best performance on systems where the
number of computing nodes is significantly smaller than the number of
points in the wall-normal direction. The global transpose of the data, which
typically constrains DNS codes to run on machines with very large networking
bandwidth, is completely avoided. We have verified that a code based on
transpose algorithms cannot yield acceptable parallel speedups when
Fast Ethernet network cards are employed. When the 10-times-faster
Gigabit Ethernet is used, the transpose-based method yields positive
speedups but cannot compete, in absolute terms, with the present PLS
method, that is capable to guarantee high parallel efficiency also on
state-of-the-art processors connected with relatively slow Fast
Ethernet cards.

As a result, the computing effort, as well as the required memory
space, can be efficiently subdivided among a number of low-cost
computing nodes. Moreover, the distribution of data in wall-parallel
slices allows us to exploit a particular, efficient and at the same
time cost-effective connection topology, where the computing machines
are connected to each other in a ring. Getting rid of the switch is
something that should not be underestimated. When the transpose-based
code is run for a $128^3$ case on two Opteron machines connected each
other point-to-point without the HP switch, the parallel speedup
increases significantly, from 1.13 to 1.53. When a third machine is
inserted between the two computing machines, the parallel speedup
becomes 1.40. While we are not in the position to extrapolate the
relevance of this result to higher-quality switches, removing the need
for a switch altogether is certainly an improvement performance-wise.

A dedicated system can be easily built, using commodity hardware and
hence at low cost, to run a computer code based on the PLS method.
Such a system grants high availability and throughput, as well as ease
in expanding/upgrading. It is our opinion that this concept of
Personal Supercomputer can be successful, since it is a specialized
system, yet built with mass-market components, and can be fully
dedicated to a single research group or even to a single researcher,
rather than being shared among multiple users through a queueing
system. Moreover, the additional investment required to specialize
towards the PLS code a general-purpose cluster of PC is simply that
required to acquire two additional network cards for each node, plus a
few meters of network cable. This means that a cluster can be easily
built in such a way that it works optimally both as a parallel
computing server for the general public and a DNS computer for a
research group employing a PLS-based code.

Concerning the (theoretical) peak computing power, we have estimated
in \cite{quadrio-luchini-floryan-2003} that the investment (early
2003) required to obtain 200 GFlop/s of peak power with a
state-of-the-art supercomputer would be 50-100 times higher than that
needed to build a Personal Supercomputer of the same power. The
smaller investment, together with additional advantages like reduced
power consumption and heat production, minimal floor space occupation,
etc, allows the user to have dedicated access to the machine for
unlimited time, thus achieving the highest throughput. As an example,
in \cite{quadrio-ricco-2004} we have performed with the Pentium
III-based machine a large number of turbulent channel flow
simulations, whose global computational demand is estimated to be
300-400 times larger than the DNS described in
\cite{kim-moin-moser-1987}.  Even though our performance measurements
have been partly carried out on relatively old computing hardware, we
have demonstrated that very good parallel speedups can be obtained for
problems whose computational size fits that of typical DNS problems
affordable with the given hardware.

The sole significant difference performance-wise between such a system
and a real supercomputer lies in the networking hardware, which offers
significantly larger bandwidth and better latency characteristics in
the latter case. However the negative effects of this difference are
not felt when the present parallel algorithm is employed, since the
need for a large amount of communication is removed {\em a priori},
thanks to the algorithm itself.

\section*{Acknowledgments}

The test of the PLS method on the Itanium II cluster has been carried
out at the SHARCNET Computing Centre at the University of Western
Ontario, Canada. Preliminary versions of this work have been presented
by M.Q. at the XV AIMETA Conference on Theoretical Mechanics
\cite{quadrio-luchini-2001} and at the XI Conference of the 
CFD Society of Canada \cite{quadrio-luchini-floryan-2003}.

\bibliographystyle{elsart-num}

\end{document}